\documentclass[aps,prl,showpacs,preprintnumbers,nofootinbib,float,longbibliography,superscriptaddress,notitlepage, twocolumn]{revtex4-2}

\usepackage[utf8]{inputenc}
\usepackage{epsfig}
\usepackage{float,amsmath,amssymb}
\usepackage{graphicx,yfonts}
\usepackage{url}
\usepackage{bbold}
\usepackage{physics}
\usepackage{enumitem}

\usepackage[normalem]{ulem}

\usepackage[]{mdframed}
\usepackage[breaklinks,colorlinks,citecolor=citcolor,urlcolor=blue,linkcolor=lcolor]{hyperref}
\definecolor{lcolor}{rgb}{0.5,0,0}
\definecolor{citcolor}{rgb}{0,0.3,0.0}

\begin{document}

\title{Nuclear collectivity and the harmonic spectrum of two-body correlations}

\author{Jean-Paul Blaizot}
\email{jean-paul.blaizot@ipht.fr}
\affiliation{Institut de Physique Th\'eorique, Universit\'e Paris Saclay, CEA, CNRS, F-91191 Gif-sur-Yvette, France}

\author{Giuliano Giacalone}
\email{giuliano.giacalone@cern.ch}
\affiliation{Theoretical Physics Department, CERN, 1211 Geneva 23, Switzerland}

\author{Alessandro Lovato}
\email{lovato@anl.gov}
\affiliation{INFN-TIFPA Trento Institute of Fundamental Physics and Applications, Via Sommarive 14, I-38123 Povo, Trento, Italy}
\affiliation{Physics Division, Argonne National Laboratory, Argonne, Illinois 60439, USA}
\affiliation{Computational Science Division, Argonne National Laboratory, Argonne, Illinois 60439, USA}

\begin{abstract}
%% BACKGROUND
High-energy nuclear collisions have opened a new experimental method to reveal collective behavior in nuclear ground states through the lens of many-body correlations of nucleons.
%% OBJECTIVES/METHODS
Using \textit{ab initio} lattice and variational calculations of $^{20}$Ne and $^{16}$O, we study how emergent phenomena such as deformation or clustering  can be identified in these systems from the dependence of their two-body density distributions on the relative azimuthal angle of nucleon pairs.
%% RESULTS
A harmonic analysis of the correlation functions reveals in particular a dominant quadrupole component in $^{20}$Ne, consistent with a \textit{bowling-pin} picture, and a prominent triangular modulation in $^{16}$O, possibly indicative of alpha-cluster correlations.
%% CONCLUSIONS
Given that such structures can be accurately identified in high-energy collider experiments, these findings open a new paradigm for analyzing emergent collective behavior in atomic nuclei, relating their intrinsic shapes to the harmonic spectrum of microscopic correlations.
\end{abstract}

\preprint{CERN-TH-2025-265}

\maketitle

%%% begin Introduction
One of the most striking features of atomic nuclei is that they exhibit emergent collective phenomena across a wide range of scales. Nuclear collectivity is traditionally inferred from patterns observed in energy spectra, which have a natural interpretation in terms of ``intrinsic states'', characterized by ``shapes'' such as quadrupole deformations or regular arrangements of alpha particles. Physical states with quantum numbers consistent with those observed in the excitation spectra are subsequently obtained from the intrinsic states via projection techniques. A recent review by Verney \cite{Verney:2025efj} highlights the immense progress in this field, from the pioneering work of Bohr and Mottelson in the 1950s to the present day.

In this work, we depart from this conventional approach, whereby the nuclear shape is primarily deduced from nuclear excitations, and focus solely on the exact ground state. We show that the intrinsic shapes can be directly identified in the harmonic decomposition of two-body correlation functions in the exact ground states, which are available thanks to recent advances in \textit{ab initio} \cite{Hergert:2020bxy,Ekstrom:2022yea} calculations of deformed nuclei. We argue, then, that the observed angular correlations of nucleons, and the Fourier harmonics associated with their harmonic decomposition are naturally amenable to a classical interpretation based on the notion of intrinsic shape.

Our approach is complementary to the traditional one. The ``vertical route'' indeed climbs the ladder of spectroscopic levels, inferring collectivity from patterns in the excitation spectrum. The ``horizontal route'' that we follow in this paper considers only the ground state, and infers emergent collective behavior from long-range angular correlations of nucleons. However, while spectroscopic techniques for studying energy spectra have long been established in nuclear experiments, accessing ground-state correlations requires novel tools. For instance, in tabletop experiments with $^6$Li atoms, breakthroughs in \textit{imaging} protocols have allowed to directly measure ground-state many-body densities, particularly two- and three-body correlations \cite{Holten:2020jrt,Holten:2021pex,Xiang:2024isi,Yao:2024rew,deJongh:2024pmo}. Such protocols are based on experimental setups capable (i) of recording atom positions in individual measurements, and (ii) of repeating such measurements a statistically significant number of times to enable the extraction of inter-atom correlations. 

An equivalent imaging at nuclear scales is provided by high-energy nucleus-nucleus collisions. Such collisions are nearly instantaneous, with the two nuclei crossing each other on times scales on the order of $10^{-26}$\,s at the CERN Large Hadron Collider (LHC). During that short time, the positions of the nucleons are frozen in the plane perpendicular to the collisions axis, granting access to their event-by-event distributions. An experimental probe is required then to expose inter-nucleon correlations. Recent work has shown that this can be achieved through measurements of azimuthal correlations of produced hadrons \cite{Jia:2022ozr,Giacalone:2023hwk,Ollitrault:2023wjk,Duguet:2025hwi,Giacalone:2025vxa}. Thus, high-energy nuclear experiments can give access to angular correlations present in the ground states of nuclei. Exhibiting such correlations on representative nuclei is the aim of this study.

%%% end Introduction

We start from nucleon coordinates in three dimensions obtained by sampling the ground-state wave functions either from continuum quantum Monte Carlo (QMC) methods~\cite{Carlson:2014vla,Gandolfi:2020pbj} or from Nuclear Lattice Effective Field Theory (NLEFT) \cite{Lee:2025req} simulations. These calculations are described in the Supplemental Material. Both are \textit{ab initio} methods, whereby many-body correlations of nucleons are generated without explicitly invoking any intermediate concepts such as clustering or deformation. We consider $J=0$ ground states characterized by a spherical one-body density and focus on two nuclei.

The first one is $^{16}$O, a doubly-magic nucleus which is spherical in mean-field approaches, yet well known for exhibiting $\alpha$-clustering effects \cite{Giacalone:2024luz}, reflecting features of a tetrahedral structure in its energy spectrum \cite{Bijker:2016bpb}. For $^{16}$O, we consider both NLEFT and QMC simulations.

The second nucleus, $^{20}$Ne, contrasts sharply with $^{16}$O. It is a doubly open-shell nucleus that displays a ground-state rotational band, signaling quadrupole deformation and strong collectivity \cite{Frosini:2021sxj}. Indeed, $^{20}$Ne has the largest ground-state deformation among stable nuclei, as inferred from spectroscopic data \cite{Pritychenko:2013gwa}. For this nucleus, we only have NLEFT ground-state configurations.

\begin{figure}
    \centering
    \includegraphics[width=.9\linewidth]{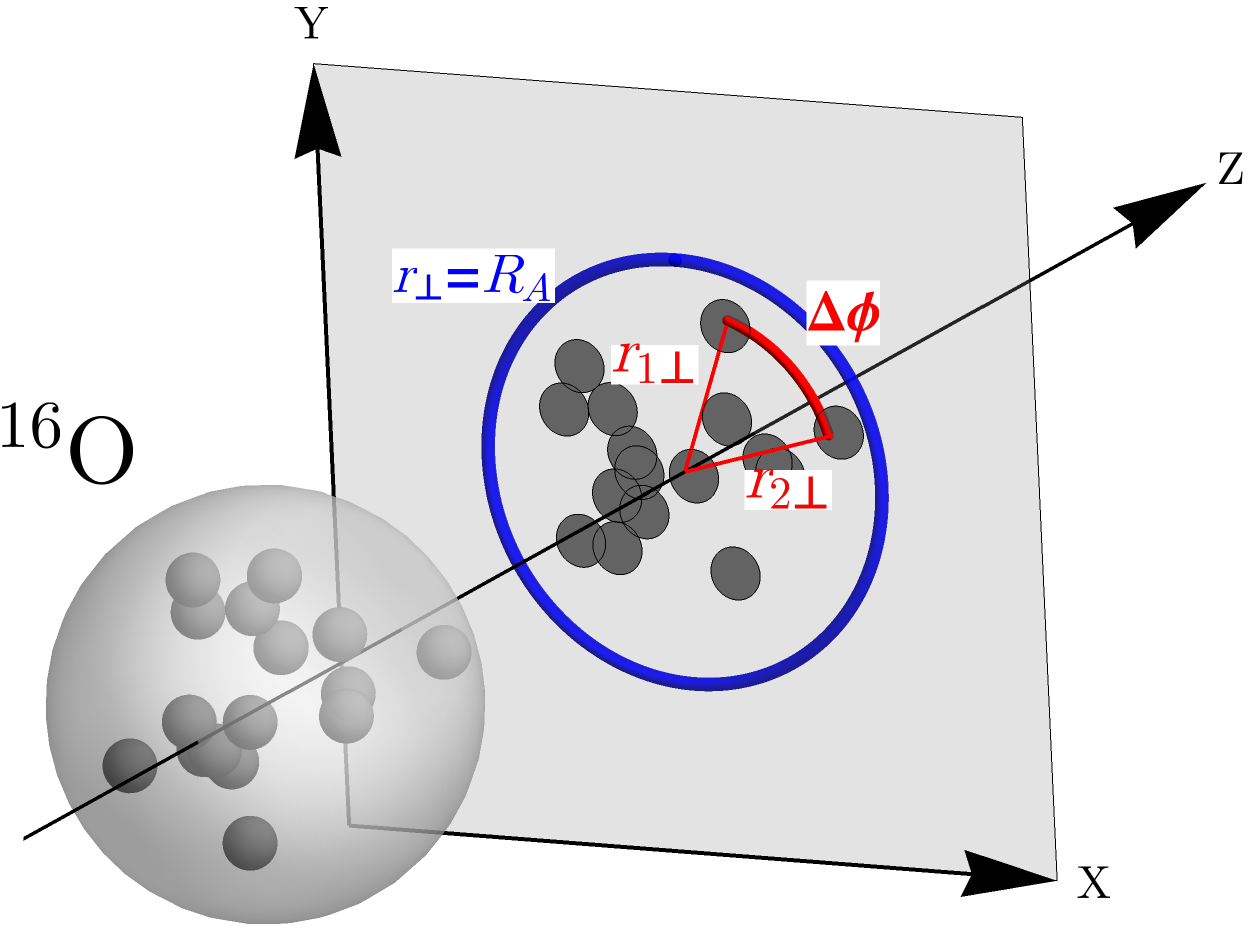}
    \caption{$^{16}$O nucleus in its rest frame (left). Probed in the laboratory frame following an acceleration to ultrarelativistic energies, the nucleus appears flattened in the $(x,y)$ plane (right), leading to a marginalization of the $z$ coordinates of its nucleons (the beam direction). In this work we are concerned with the two-body density $\rho^{(2)}(r_{1\perp}=r_\perp,r_{2\perp}=r_\perp,\Delta\phi=\phi_1-\phi_2)$, obtained from the projection of a large number of nuclei (snapshots), and how it varies as a function of the relative azimuthal angle, $\Delta\phi$, in a circle of radius $r_\perp$.}
    \label{fig:1}
\end{figure}

With these \textit{ab initio} results in hand, we calculate the one- and two-body densities for each system, with 
\begin{align}
\rho^{(1)} ({\bf r}) & = \langle \Psi | \sum_{i}^A \delta(\mathbf{r}_1 - \hat{\mathbf{r}}_i) | \Psi \rangle \\
\label{eq:rho12} \rho^{(2)}(\mathbf{r}_1,\mathbf{r}_2) &= \langle \Psi | \sum_{i \neq j = 1}^A \delta(\mathbf{r}_1 - \hat{\mathbf{r}}_i) \delta(\mathbf{r}_2 - \hat{\mathbf{r}}_j) | \Psi \rangle,
\end{align}
where $\textbf{r}=(r_x,r_y,r_z)$, $A$ is the number of nucleons, and spin and isospin degrees of freedom are averaged out. The densities satisfy the sum rules
\begin{align}
    \int_{{\bf r}_1} \rho^{(2)}(\mathbf{r}_1,\mathbf{r})= (A-1) \rho^{(1)}({\bf r}), \hspace{10pt}
    \int_{{\bf r}} \rho^{(1)}(\mathbf{r})= A \,.
\end{align}
As illustrated in Fig.~\ref{fig:1}, in high-energy nuclear scattering experiments the probed densities are projected on the transverse plane, that is, the plane perpendicular to the collision axis, chosen to be the $z$-axis. We then define
\begin{equation}
\rho^{(2)}_\perp(r_{1x},r_{1y},r_{2x},r_{2y}) = \int dr_{1z} dr_{2z} \, \rho^{(2)}(\mathbf{r}_1,\mathbf{r}_2) \, ,
\end{equation}
which depends on four transverse coordinates. The dynamics in the longitudinal ($z$) direction is blurred by particle production mechanisms \cite{Mantysaari:2024qmt}. However these processes are local  in the transverse plane, and do not alter significantly the long range correlation effects that we are after, which are mostly determined by the transverse positions of the nucleons and captured by the function $\rho^{(2)}_\perp$.

The recent analysis of Ref.~\cite{Blaizot:2025scr} based on a classical rigid-rotor picture demonstrates that, for small deformations relevant for the ground states of stable isotopes, the correlated part of the two-body density acquires the form 
\begin{align}
\nonumber \rho^{(2)}_\perp ({\bf r}_{1\perp}, {\bf r}_{2\perp}) - \rho^{(1)}({\bf r}_{1\perp}) & \rho^{(1)}({\bf r}_{2\perp}) \\ 
&\propto \delta_n^2 \, r_{1\perp}^n \, r_{2\perp}^n \, \cos(n\Delta\phi), 
\label{eq:blaizot}
\end{align}
where  ${\bf r}_{\perp}  = (r_x, r_y) = (r_\perp, \phi)$ and $\Delta\phi=\phi_1-\phi_2$ (as shown in Fig.~\ref{fig:1}), and where $\delta_n$ is a measure of the deformation of the intrinsic shape of the nucleus in the $n$th order Fourier harmonic. Our main objective is to identify this behavior in the harmonic analysis of the two-body densities of the nuclei $^{16}$O and $^{20}$Ne obtained from the \textit{ab initio} NLEFT and QMC calculations.

\begin{figure*}[t]
    \centering
    \includegraphics[width=\linewidth]{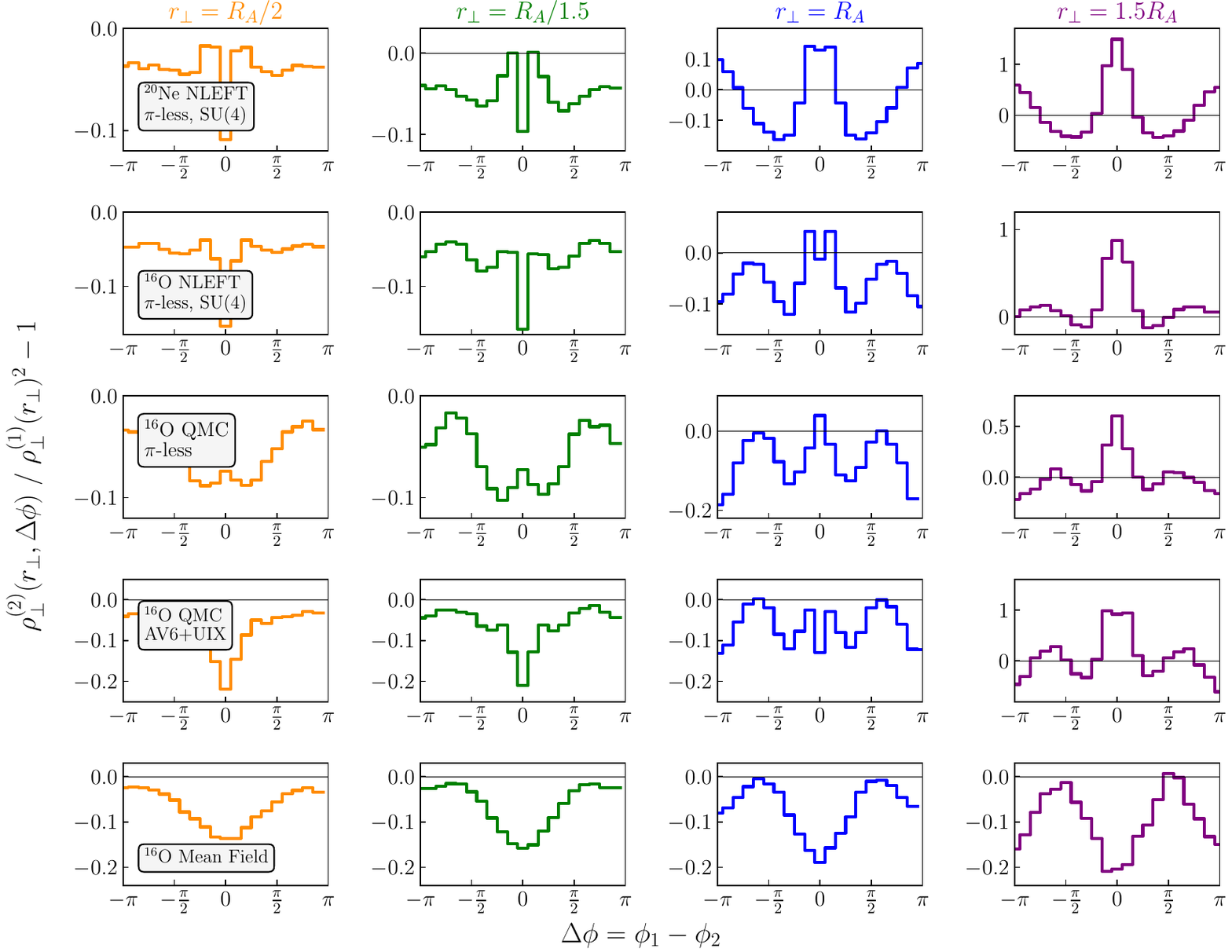}
    \caption{Emergent collective behavior of nucleons in the ground states of $^{20}$Ne (upper row) $^{16}$O (second, third, and fourth row). The two-body azimuthal correlation is plotted as a function of $\Delta\phi = \phi_1 - \phi_2$, for radial slices $r_\perp = R_A/2$ (first column), $R_A/1.5$ (second column) $R_A$ (third column), and $1.5 R_A$ (fourth column). Different rows indicate different \textit{ab initio} nuclear structure input: NLEFT configurations with a soft SU(4)-symmetric potential are for $^{20}$Ne and $^{16}$O (first and second row, respectively); QMC calculations employing either a pion-less EFT Hamiltonian (third row), or the phenomenological AV6P+UIX potential (fourth row). For $^{20}$Ne, a dominant $\cos(2\Delta\phi)$ quadrupole modulation emerges as one approaches $r_\perp=R_A$, consistent with a quadrupole-deformed rotor picture. In $^{16}$O, a $\cos(3\Delta\phi)$ modulation emerges as we approach the nuclear surface ($r_\perp \sim R_A$), indicating an effect akin to an octupole deformation. The observed dips at $\Delta\phi=0$ for small $r_\perp$ are a consequence of short-range repulsion among nucleons. This is showcased by the results in the fifth row, obtained for a mean-field configuration (Slater determinant) of $^{16}$O, where correlations are solely induced by Pauli exclusion.
    \label{fig:2}}
\end{figure*}

%%% RESULTS
Our results are shown in Fig.~\ref{fig:2}. To simplify both the discussion and the computations, we set $r_{1\perp} = r_{2\perp} = r_\perp$ in the two-body densities and consider representative values of $r_\perp$ given as fractions of $R_A$, which is the rms nuclear radius in three dimensions (available in Tab.~\ref{tab:1}). Increasing $r_\perp$ from the center to the periphery of the nucleus corresponds to moving from left to right in Fig.~\ref{fig:2}. Moving vertically, different isotopes, different many-body methods for solving the Schr\"odinger equation, as well as different underlying nuclear Hamiltonians are considered. The panels show the normalized correlation function
\begin{equation}
    \rho^{(2)}_\perp (r_\perp,\Delta\phi) \, / \,  \rho^{(1)}_\perp (r_\perp)^2 - 1 \, ,
\end{equation}
as a function of the relative azimuthal angle. This function reduces to $-1/A$ in absence of correlations, and, as discussed in the Supplemental Material, is in general negative-definite for a system of non-interacting fermions (where correlations arise from Pauli exclusion). As also explained in the Supplemental Material, we employ coordinate configurations which do not include any recentering effects, revealing thus truly dynamical correlations encoded in the nuclear forces.

The first row of Fig.~\ref{fig:2} shows results for the nucleus $^{20}$Ne, available from the NLEFT simulations employing a low-resolution Hamiltonian based on the Wigner SU(4) symmetry \cite{Lu:2018bat}. We see that, as one moves in $r_\perp$ from the center to the periphery of the nucleus, the two-body correlation acquires a $\cos(2\Delta\phi)$ modulation, which grows in amplitude as one approaches the nuclear surface. This fully validates the expectation of Eq.~(\ref{eq:blaizot}), i.e., a harmonic spectrum consistent with the well-deformed rotor interpretation that is expected to hold for the nucleus $^{20}$Ne.

Results for $^{16}$O are shown in the second, third, and fourth rows of Fig.~\ref{fig:2}. Respectively, they show results from the NLEFT calculations, from the QMC calculations implementing a soft pion-less EFT Hamiltonian, and from the QMC calculations with a phenomenological AV6P+UIX potential (see the Supplemental Material for further descriptions). As we move towards $r_\perp=R_A$, we observe the emergence of a striking pattern: a clean $\cos(3\Delta\phi)$ modulation, consistently obtained for all many-body methods and Hamiltonians. Following Eq.~(\ref{eq:blaizot}), this suggests the presence of octupole correlations in the ground state of $^{16}$O, in agreement with an alpha-clustered shape for this nucleus \cite{Giacalone:2024luz}: projecting a tetrahedron onto $(x,y)$ leads to a triangular shape whose triangularity depends on how strongly nucleons are clustered \cite{YuanyuanWang:2024sgp}.

%% Comment on dips and short-range correlations
All  the curves shown in Fig.~\ref{fig:2} exhibit dips at small $r_\perp$ and for $\Delta\phi = 0$. One may intuitively relate these dips to  short-range repulsive effects. This is particularly manifest in the lowermost row of results in Fig.~\ref{fig:2}, where the oxygen configurations are sampled from a mean-field state (Slater determinant) in which correlations, particularly the observed dips at $\Delta\phi=0$, arise uniquely as a consequence of Pauli exclusion effects. These correlations appear to dominate for low values of $r_\perp$, explaining why all our curves in Fig.~\ref{fig:2} are negative in such a regime. For the NLEFT results, the sharp dips at small $r_\perp$ are due to a minimum inter-nucleon distance of 0.5 fm (irrespective of spin or isospin), which is imposed by hand in the sampled coordinates~\cite{Giacalone:2024luz}. In contrast, the QMC results feature repulsive effects encoded in the Hamiltonians \cite{Zhang:2024vkh,Liu:2025uks}.  We also note that, for the largest $r_\perp$, all curves except the mean-field result exhibit a strong correlation peak at $\Delta\phi = 0$. This points to an interaction-driven pairing of nucleons in the periphery of the nuclei, reminiscent of pairing phenomena recently reported in dilute gases of strongly-interacting ultra-cold fermions \cite{Brandstetter:2023jsy,Brandstetter:2024gur,Yao:2024rew},  warranting further investigation.

\begin{figure}
    \centering
    \includegraphics[width=0.75\linewidth]{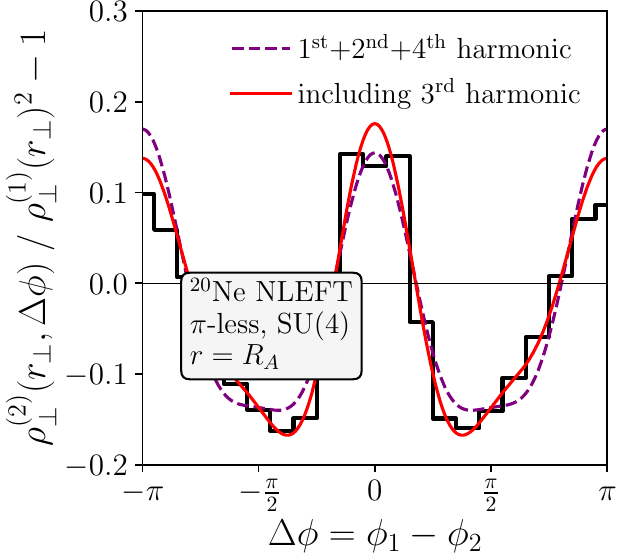}
    \caption{Fourier series decomposition of the two-body correlation of $^{20}$Ne as a function of $\Delta\phi$ for $r=R_A$. The solid black histogram corresponds to the curve shown in the upper row of Fig.~\ref{fig:2}. The  decomposition is of the form $a_0+\sum_{n=1}^4 a_n \cos(n\Delta\phi)$, with $a_0=-0.0289$, $a_1=-0.0130$, $a_2=0.147$, $a_3=0.0322$, $a_4=0.0389$. The purple dashed line corresponds to the Fourier series including only $a_1$, $a_2$, and $a_4$. The red solid line shows the effect of including $a_3$, which notably shifts the minima of the curve to their true location. }
    \label{fig:3}
\end{figure}

%% Here discuss <En> operators

In analogy with the analysis of collective flows in heavy-ion collisions, we characterize the correlation functions displayed in Fig.~\ref{fig:2} in terms of appropriate moments, or generalized eccentricities  \cite{Blaizot:2014nia}. 
Following Ref.~\cite{Duguet:2025hwi}, we consider  the following two-body operator
\begin{align}
\label{eq:Ehat}
\hat{\mathcal{E}}_n ({\bf r}_1, {\bf r}_2) &= r_{1\perp}^n \, r_{2\perp}^n \, e^{i n (\phi_1-\phi_2)},
\end{align}
whose expectation value quantifies two-body azimuthal correlations in the transverse plane. The ground state expectation value of such a local operator has a natural expression in terms of the density-density correlation function $S(\mathbf{r}_1,\mathbf{r}_2)=\rho^{(1)}(\mathbf{r}_1)\delta(\mathbf{r}_1,\mathbf{r}_2)+\rho^{(2)}_c(\mathbf{r}_1,  \mathbf{r}_2)$ where $\rho^{(2)}_c(\mathbf{r}_1,  \mathbf{r}_2)= \rho^{(2)}(\mathbf{r}_1,  \mathbf{r}_2)-\rho^{(1)}(\mathbf{r}_1)\rho^{(1)}(\mathbf{r}_2)$ is the connected part of $\rho^{(2)}$. However, $S(\mathbf{r}_1,\mathbf{r}_2)$  contains a (sizable) local contact term which does not carry direct information on two-body correlations \cite{Blaizot:2014nia}. Therefore, we define
\begin{equation}\label{eq:normaEhat}
    \langle \hat{\mathcal{E}}_n\rangle = \frac{1}{A(A-1)} \int_{{\bf r}_1, {\bf r}_2} \rho^{(2)}({\bf r}_1, {\bf r}_2) \, r_{1\perp}^n \, r_{2\perp}^n \, e^{i n (\phi_1-\phi_2)} \,
\end{equation}
keeping only the two-body density in the expectation value in order to quantify the impact of correlations.

Since $\langle \hat{\mathcal{E}}_n \rangle$ is naturally proportional to $R^{2n}$, where $R$ is the typical transverse size of the nucleus, we quantify the deformation of the two-body density via the dimensionless parameter
\begin{equation}
\mathfrak{B}_{n}^{2} = \frac{2 \pi}{3} \frac{\langle   \hat{\mathcal{E}}_{n}   \rangle}{\langle   \hat{R}_{n}   \rangle^2}, \hspace{10pt} \langle \hat{R}_n \rangle = \frac{1}{A}\int_{\bf r} (r_{\perp}^2)^{n/2} \rho^{(1)} (r) \,.
\end{equation}
The prefactor $2\pi/3$ is motivated by the analysis of Ref.~\cite{Duguet:2025hwi} for a classical rigid-rotor picture in which $\rho^{(2)}({\bf r}_1, {\bf r}_2)$ in the lab frame is given by the orientation average of an intrinsic density of nucleons, having a deformed surface described via the standard deformation parameters $\beta_n$. For a deformed Gaussian density with small deformations, $\mathfrak{B}_n$ is indeed closely related to $\beta_n$, with $\mathfrak{B}_2 = \beta_2$ and $\mathfrak{B}_3 = 1.06 \, \beta_3$. These relations should hold to high accuracy beyond the Gaussian approximation, e.g., for more realistic liquid-drop or Woods-Saxon distributions \cite{Jia:2021tzt,Jia:2021qyu}. Our results for the moments of the analyzed configurations are provided in Tab.~\ref{tab:1}.

\begin{table*}[t]
\centering
\begin{tabular}{l l l c c c c c c c}
\hline
Species & Method & Interaction & $R_A$\,[fm] &
$\langle \hat{R}_2 \rangle$\,[fm$^{2}$] &
$\langle \hat{R}_3 \rangle$\,[fm$^{3}$] &
$\langle \hat{\mathcal{E}}_2 \rangle$\,[fm$^{4}$] &
$\mathfrak{B}_2$ &
$\langle \hat{\mathcal{E}}_3 \rangle$\,[fm$^{6}$] &
$\mathfrak{B}_3$ \\
\hline

% Mean Field 16O
$^{16}\mathrm{O}$ & Mean Field [Woods-Saxon] & / & 2.76 &
5.09 & 15.0 & -0.763 & -0.249 & $\approx 0$ & $\approx 0$ \\

% QMC 16O (pi-less EFT)
$^{16}\mathrm{O}$ & QMC & $\pi$-less EFT & 2.48 &
4.12 & 10.7 & -0.230 & -0.169 & 5.11 & 0.305 \\

% QMC 16O (AV6P+UIX)
$^{16}\mathrm{O}$ & QMC & AV6P+UIX & 2.62 &
4.60 & 12.0 & -0.125 & -0.111 & 14.9 & 0.465 \\

% NLEFT 16O
$^{16}\mathrm{O}$ & NLEFT & SU(4) & 2.78 &
5.18 & 15.3 & 0.521 & 0.202 & 13.5 & 0.347 \\

% NLEFT 20Ne
$^{20}\mathrm{Ne}$ & NLEFT & SU(4) & 3.17 &
6.74 & 22.8 & 4.39 & 0.450 & 25.2 & 0.318 \\

\hline
\end{tabular}
\caption{Properties of the ground-state configurations of nucleons.}
\label{tab:1}
\end{table*}

Starting with the NLEFT results for $^{20}$Ne, we find both large quadrupole and octupole deformation coefficients. This points to the \textit{bowling-pin}, or $^{16}$O+$\alpha$ shape advocated in Ref.~\cite{Giacalone:2024luz}, which is strongly elongated along the nuclear axis (large quadrupole parameter $\beta_2\approx\mathfrak{B}_2\approx0.5$), as well as strongly reflection-asymmetric (large $\beta_3\approx\mathfrak{B}_3\approx0.3$). Relating to the curves in Fig.~\ref{fig:2}, we clarify this point by performing a Fourier decomposition of the correlation function of $^{20}$Ne for $r_\perp=R_A$. The result is shown in Fig.~\ref{fig:3}. The effect of the third harmonic is to visibly skew the curve at its minima, determining a shift in their location from the value $\Delta\phi=\pm\pi/2$ expected for a quadrupole modulation alone. As we mentioned, the leading-order expressions of the mean squared eccentricities, $\langle \varepsilon_n^2 \rangle$, relevant for the collective flow analyses in high-energy collision experiments, are \cite{Duguet:2025hwi}
\begin{equation}
    \langle \varepsilon_n^2 \rangle = \frac{1}{2A \, \langle \hat{R}_n \rangle^2} \left [ \langle \hat{R}_{2n} \rangle + (A-1) \langle \hat{\mathcal{E}}_n \rangle \right ] \, .
\end{equation}
Setting $n=3$, for $^{20}$Ne we obtain $\langle \hat{R}_{6} \rangle / \langle \hat{R}_{3} \rangle^2 = 3$, while $(A-1)\langle \hat{\mathcal{E}}_3 \rangle/\langle \hat{R}_{3} \rangle^2=0.94$. Therefore, the mean squared triangularity receives a sizable 30\% correction from genuine two-body correlations, that is, the distortion of the minima in Fig.~\ref{fig:3}. Since final-state anisotropies from the recent LHC run of $^{20}$Ne ion collisions are measured with extremely small uncertainties \cite{ATLAS:2025nnt,ALICE:2025luc,CMS:2025tga}, we expect that the imprint of the \textit{bowling-pin} structure of neon should be fully resolved in the current data sets.

Moving to $^{16}$O in Tab.~\ref{tab:1}, our first remark is that, for the QMC configurations, the quadrupole correlation is negative, $\langle \hat{\mathcal{E}}_2 \rangle<0$. Intuitively, this is a consequence of the central dips in the correlation functions discussed in Fig.~\ref{fig:2}. Indeed, a dip at $\Delta\phi=0$ followed by peaks around $\Delta\phi=\pm\pi/2$ represents a correlation structure consistent with a negative Fourier coefficient in the second harmonic. This is especially clear in the mean-field calculations of $^{16}$O (last line in Fig.~\ref{fig:2} and first line in Tab.~\ref{tab:1}). Pauli exclusion leads to a hole around $\Delta\phi=0$, and as $r_\perp$ increases, a modulation of the form $-\cos(2\Delta\phi)$ emerges, in turn giving $\langle \hat{\mathcal{E}}_2 \rangle <0$. This behavior is elucidated through analytical computations in the Supplemental Material. While these repulsive effects appear to be overcome in the $^{16}$O NLEFT configurations, presenting $\mathfrak{B}_2>0$, in the QMC results the oxygen nucleus does not develop enough quadrupole collectivity to flip the sign of $ \langle \hat{\mathcal{E}}_2 \rangle$, which remains negative, although closer to zero than for the mean-field baseline. 

That said, in all considered models the oxygen nucleus develops strong octupole correlations that lead to a large and positive $\mathfrak{B}_3$ parameter, with the mean-field baseline giving $\mathfrak{B}_3=0$. This is consistent with the curves of Fig.~\ref{fig:2} and with an alpha-clustered shape for this isotope, which we directly probe through the Fourier decomposition.

%%Conclusion
In conclusion, collectivity in atomic nuclei can be observed in the form of emergent harmonic modulations of the ground-state two-body densities, which have an immediate interpretation in terms of deformed intrinsic shapes. These patterns are of experimental relevance because they impact the final-states of high-energy nuclear collisions. Measurements of elliptic and triangular flows in the limit of ultra-central collisions are indeed to leading order sensitive to the correlation structures observed in Fig.~\ref{fig:2}. Particularly, experimental data from LHC runs on O+O and Ne+Ne collisions \cite{ATLAS:2025nnt,ALICE:2025luc,CMS:2025tga} already reveal non-trivial relative azimuthal anisotropies in the comparisons of the two systems, which could be used to infer the values of $\mathfrak{B}_n$ and detailed many-body correlations.

%% Outlook
Clearly, this is only the tip of the iceberg. Our study could be generalized to three-body correlations probing the \textit{triaxiality} of the ground states, also accessible in collider experiments \cite{Bally:2021qys,Jia:2021qyu,ATLAS:2022dov,Zhao:2024lpc,Hagino:2025vxe}. Furthermore, leveraging recent advances in statistical methods and emulators \cite{Sun:2024iht}, it becomes possible to investigate how the emergence of the harmonic spectrum of nuclei is driven by the low-energy constants of chiral EFT expansions \cite{Piarulli:2019cqu,Hammer:2019poc}. This would elucidate how collider data probes the nuclear force, as seen as an EFT of low-energy QCD. Ultimately, this new experimental information could be used to perform sophisticated benchmarks of \textit{ab initio} methods of nuclear structure for deformed species. This could in turn help address fundamental open issues, such as the  determination of nuclear matrix elements of neutrinoless double beta decay \cite{Yao:2021wst,Agostini:2022zub}, which are sensitive to the ground-state many-body correlations discussed in this work \cite{Li:2025vdp}.

\bigskip
 We thank Dean Lee, Shihang Shen, and all the members of the Nuclear Lattice Effective Field Theory collaboration for allowing us to show results from pinhole configurations derived from their method. We thank the participants of the ESNT - CEA Saclay workshop \textit{``Charge and matter distributions in nuclei''} for useful discussions.

\section{Supplemental Material}

\subsection{\textit{Ab initio} nuclear input: QMC}

The quantum Monte Carlo (QMC) calculations discussed in this work are based on nonrelativistic Hamiltonians of the form
\begin{equation}
H = - \sum_{i} \frac{\nabla_i^2}{2m_N} + \sum_{i<j}v_{ij} + \sum_{i<j<k} V_{ijk}, ,
\label{eq:hamiltonian}
\end{equation}
where $v_{ij}$ and $V_{ijk}$ denote the nucleon--nucleon ($NN$) and three-nucleon ($3N$) interactions, respectively. Specifically, we employ two complementary Hamiltonians that differ substantially in their resolution.
The first Hamiltonian is low resolution, as it is based on a pionless effective field theory expansion—specifically, model o'' of Ref.~\cite{Schiavilla:2021dun}. This $NN$ potential is designed to reproduce the $np$ scattering lengths and effective ranges in the $S/T = 0/1$ and $1/0$ channels. The authors of Ref.~\cite{Schiavilla:2021dun,Gnech:2023prs} explored several regulator values for the $3N$ interaction; here we adopt $R_3 = 1.1$~fm because, when used in conjunction with model o'', this choice yields binding energies in reasonably good agreement with experiment for both $^4$He and $^{16}$O.
The second Hamiltonian is high resolution and consists of the Argonne $v_6^\prime$ $NN$ potential~\cite{Wiringa:2002ja} complemented by the Urbana IX $3N$ force~\cite{Pudliner:1995wk}. For both Hamiltonians, we include only the Coulomb repulsion between finite-size protons as the electromagnetic component of the $NN$ interaction.

The Auxiliary Field Diffusion Monte Carlo (AFDMC) method~\cite{Schmidt:1999lik,Carlson:2014vla, Gandolfi:2020pbj}
employs imaginary-time projection to extract the ground state of a the nuclear
Hamiltonian starting from a suitably chosen variational wave function with
nonvanishing overlap with the exact ground state
\begin{equation}
|\Psi_0\rangle
= \lim_{\tau \to \infty} |\Psi_\tau\rangle\quad,\quad |\Psi_\tau\rangle=e^{-(H - E_V)\tau}\,|\Psi_V\rangle \,.
\end{equation}
In the above equation, \(E_V\) is an energy offset introduced to control the normalization of the
propagated wave function and is typically chosen close to the exact ground-state
energy \(E_0\).

The variational wave function is written as the product of a correlation operator
and an antisymmetric mean-field state~\cite{Lonardoni:2017hgs}
\begin{equation}
|\Psi_V\rangle
= \left(F_c + F_{2b} + F_{3b}\right)\,|\Phi\rangle_{J^\pi M T_z} .
\end{equation}
Here, \(F_c\) accounts for spin- and isospin-independent two- and three-body
correlations. To keep the computational cost polynomial in the number of
nucleons \(A\), the operators \(F_{2b}\) and \(F_{3b}\) include linearized
spin- and isospin-dependent two- and three-body correlations, as described in
Ref.~\cite{Gandolfi:2014ewa}.

The long-range component \( |\Phi\rangle \) is taken to be a shell-model-like state with good total angular momentum \(J\), its projection \(M\), parity \(\pi\), and isospin projection \(T_z\). For the closed-shell \(^{16}\)O nucleus considered in this work, a single Slater determinant is sufficient to reproduce the correct quantum numbers,
\begin{equation}
|\Phi\rangle_{J^\pi M T_z}
=
\mathcal{A}
\big(
\phi_{\alpha_1}^{(n)} \cdots \phi_{\alpha_A}^{(n)}
\big)_{J^\pi M T_z} \,,
\end{equation}
where \(\mathcal{A}\) enforces antisymmetrization. The single-particle orbitals \(\phi_\alpha^{(n)}\) are obtained by solving the single-particle Schr\"odinger equation in a Woods--Saxon potential, with parameters optimized variationally.

The mean-field calculations reported in this work are obtained by setting \( F_c = 1 \) and \( F_{2b} = F_{3b} = 0 \), and by using the Seminole parameterization of the Woods--Saxon potential~\cite{Schwierz:2007ve} to generate the single-particle orbitals. The harmonic-oscillator configurations are obtained in a similar fashion, but using harmonic-oscillator single-particle orbitals, consistent with the discussion in the section below.

The ground-state expectation value of an operator \( O \) that does not commute with the Hamiltonian can be estimated in perturbation theory
as~\cite{Carlson:2014vla,Gandolfi:2020pbj}
\begin{equation}
\frac{\langle \Psi_\tau | O |\Psi_\tau\rangle}{\langle \Psi_\tau |\Psi_\tau\rangle}
\simeq
2 \frac{\langle \Psi_V | O |\Psi_\tau\rangle}{\langle \Psi_V |\Psi_\tau\rangle}
-
\frac{\langle \Psi_V | O |\Psi_V\rangle}{\langle \Psi_V |\Psi_V\rangle}\, .
\end{equation}
The two-body density falls into this category and is estimated using both Variational MC (VMC) and DMC configurations. The former correspond to statistically
independent Metropolis--Hastings samples drawn from
\begin{equation}
\pi_{\Psi_V}(R,S) =
\frac{|\Psi_V(R,S)|^2}{\sum_S \int dR\, |\Psi_V(R,S)|^2}\, ,
\end{equation}
where \( R = \mathbf{r}_1,\dots,\mathbf{r}_A \) and
\( S = s_1,\dots,s_A \) denote the spatial and spin--isospin coordinates
of the \( A \) nucleons. Since in this work we are not interested in the
spin--isospin degrees of freedom, these are marginalized, so that the
nucleon positions are distributed according to
\( \sum_S \pi_{\Psi_V}(R,S) \).

On the other hand, the DMC configurations are sampled from
\begin{equation}
\pi_{\Psi_\tau}(R,S) =
\frac{\Psi_V^*(R,S)\,\Psi_\tau(R,S)}
{\sum_S \int dR\, \Psi_V^*(R,S)\,\Psi_\tau(R,S)}\, ,
\end{equation}
where the spin--isospin degrees of freedom are again marginalized. Importantly, spurious center-of-mass contributions are automatically removed from all observables~\cite{Massella:2018xdj} by computing the
orbitals in intrinsic spatial coordinates \( \bar{\mathbf{r}}_i = \mathbf{r}_i - \mathbf{R}_{\rm CM} \), where \( \mathbf{R}_{\rm CM} \) denotes the center of mass of the nucleus.

\subsection{\textit{Ab initio} nuclear input: NLEFT}

Nuclear Lattice Effective Field Theory (NLEFT) provides a framework for solving the nuclear $A$-body problem on a finite $L \times L \times L \times N_t$ space-time lattice, where $L$ and $N_t$ denote spatial and temporal extents, respectively. Here a lattice spacing $a = 1.3155$\,fm and $L=8$ were used. The method is particularly suited for providing initial conditions of high-energy nuclear collisions, and we refer to Refs.~\cite{Giacalone:2024luz, Lee:2025req} for more extensive descriptions.
 
As in previous works, this action adopts the so-called essential elements model~\cite{Lu:2018bat}. It corresponds to a leading-order, SU(4)-symmetric effective field theory built from smeared contact operators,
\begin{equation}
H_{\rm SU(4)} = H_{\rm free}
+ \frac{1}{2!} C_2 \sum_{\mathbf{n}} \tilde{\rho}(\mathbf{n})^2
+ \frac{1}{3!} C_3 \sum_{\mathbf{n}} \tilde{\rho}(\mathbf{n})^3 ,
\end{equation}
where $\mathbf{n} = (n_x, n_y, n_z)$ labels lattice sites. The term $H_{\rm free}$ denotes the free Hamiltonian with a nucleon mass $m = 938.9$\,MeV, while the density operator $\tilde{\rho}(\mathbf{n})$ is given by~\cite{Elhatisari:2017eno}
\[
\tilde{\rho}(\mathbf{n}) =
\sum_i \tilde{a}_i^\dagger(\mathbf{n}) \tilde{a}_i(\mathbf{n})
+ s_L \sum_{|\mathbf{n}' - \mathbf{n}| = 1}
\sum_i \tilde{a}_i^\dagger(\mathbf{n}') \tilde{a}_i(\mathbf{n}') ,
\]
where $i$ represents combined spin-isospin indices. The non-locally smeared operators $\tilde{a}_i$ are defined by
\[
\tilde{a}_i(\mathbf{n}) = a_i(\mathbf{n})
+ s_{NL} \sum_{|\mathbf{n}' - \mathbf{n}| = 1} a_i(\mathbf{n}') .
\]
This theory has, thus, two low-energy constants, $C_2$ and $C_3$, which determine, respectively, the overall strength of two- and three-body interactions. The coefficients $s_L$ and $s_{NL}$ regulate the local and nonlocal smearing. Following~\cite{Lu:2018bat}, the values used are $C_2 = -3.41 \times 10^{-7}\,$MeV$^{-2}$, $C_3 = -1.4 \times 10^{-14}\,$MeV$^{-5}$, $s_{NL} = 0.5$, and $s_L = 0.061$.

In the NLEFT simulations, a trial many-body state, typically a Slater determinant, is projected to the physical ground state by Euclidean-time evolution. The so-called pinhole algorithm~\cite{Elhatisari:2017eno} enables Monte Carlo sampling of the spatial $A$-body density. It is obtained by inserting an $A$-body density operator in the projection amplitude midway through the Euclidean time evolution:
\[
Z_{f,i}^{\rm pinhole}
= \langle \Psi_f |
M^{N_t/2}
\, \rho_{i_1, \ldots, i_A}
(\vec{n}_1, \ldots, \vec{n}_A)
\, M^{N_t/2}
| \Psi_i \rangle ,
\]
where $M$ is the transfer matrix, applied $N_t$ times, and $\vec{n}$ are spatial lattice sites (``pinhole positions'') of $A$ nucleons. This method identifies the nuclear center of mass, enabling the calculation of translation-invariant matter distributions. The recorded pinhole positions carry many-body correlations to all orders. In this work, we use the positive-weight pinhole configurations of $^{16}$O and $^{20}$Ne used in the hydrodynamic calculations of Ref.~\cite{Giacalone:2024luz}.

\begin{figure*}[t]
    \centering
    \includegraphics[width=.95\linewidth]{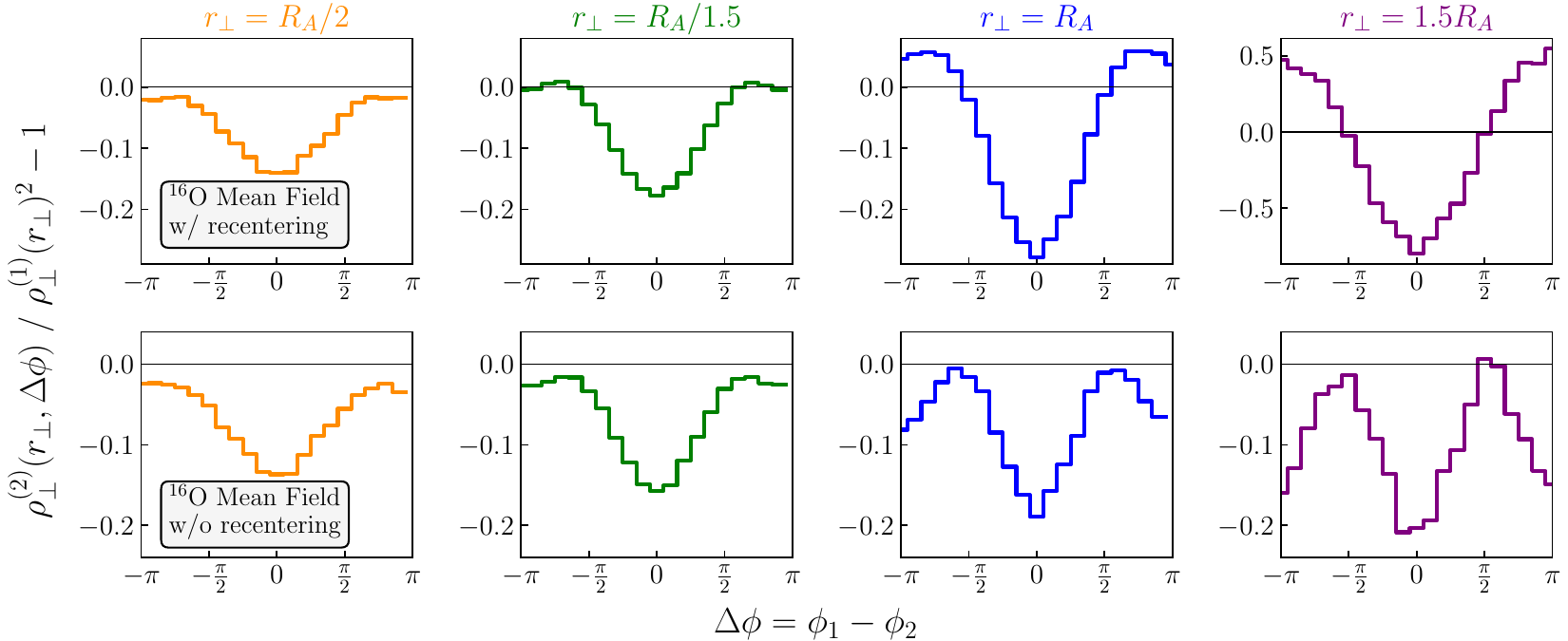}
    \caption{Angular correlation analysis for the mean field configurations of $^{16}$O. Upper: recentered configurations. Lower: with a fluctuating center-of-mass, corresponding to the fifth row in Fig.~\ref{fig:2}.}
    \label{fig:COM_MF}
\end{figure*}

\begin{figure*}[t]
    \centering
    \includegraphics[width=.95\linewidth]{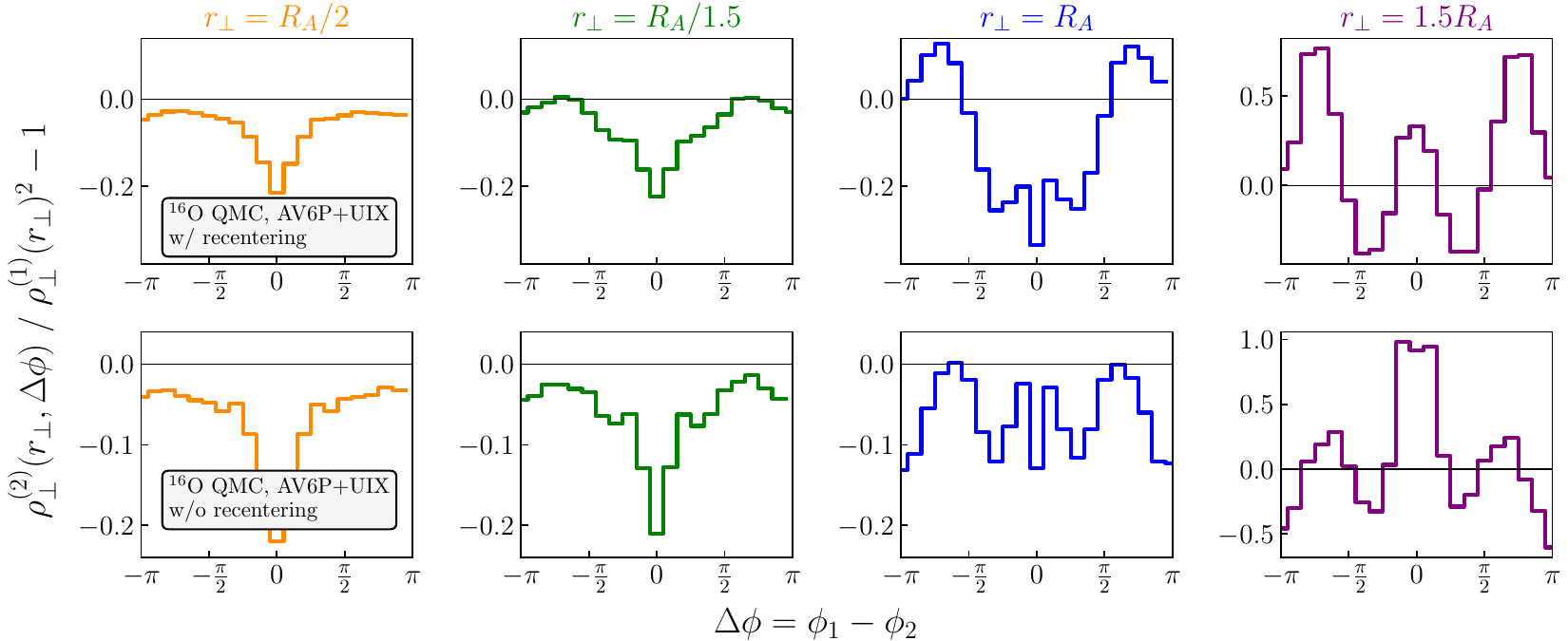}
    \caption{Same as Fig.~\ref{fig:COM_MF} but for the QMC calculations of $^{16}$O employing the AV6P+UIX potential. The lower panel corresponds to the fourth row in Fig.~\ref{fig:2}.}
    \label{fig:COM_UIX}
\end{figure*}

\subsection{\textbf{Removing recentering bias}}

By design, the configurations of nucleons returned by the QMC and NLEFT simulations are provided re-centered with respect to their own center-of-mass, that is, each configuration $\{ {\bf r}_1, \ldots, {\bf r}_A \}$ fulfills the constraint $\sum_i^A{\bf r}_i=0$. As shown, e.g., in Ref.~\cite{Zhang:2024vkh}, this recentering generates spurious correlations among configurations that will impact as well the two-body angular distributions analyzed in this article.

Therefore, in our analysis we eliminate correlations induced by this constraint and isolate the impact of dynamical many-body effects governed by the nuclear Hamiltonians.
To do so, we proceed as follows:
\begin{enumerate}
    \item We take the original re-centered configurations and calculate their one-body density, $\rho^{(1)}_C({\bf r})$.
    \item We perform a sampling of $A$ independent coordinates of nucleons in space from the distribution $\rho_C^{(1)}({\bf r})$. With the origin located at the center of the one-body density, that is, $\int_{\bf r} {\bf r}\, \rho_C({\bf r})=0$, each sampled configuration has a non-zero center-of-mass $\vec \delta_C$, whose value we store for each sampled set of $A$ nucleons.
    \item We go back to the original re-centered configurations, and shift them each by a different $\vec \delta_C$.
\end{enumerate}
This effectively returns a set of configurations that no longer carry many-body correlations associated with recentering effects.

In Ref.~\cite{Zhang:2024vkh}, the impact of the center-of-mass constraint for the two-nucleon density is discussed as a function of relative distance in a three-dimensional space. Here we display the impact of these center-of-mass effects on the two-body angular correlation studied in this manuscript. In Fig.~\ref{fig:COM_MF}, the $\Delta\phi$ dependence of the two-body density as a function of $r_\perp$ is shown for the mean-field configurations of $^{16}$O. 

The lower panel shows the results for the configurations where the recentering correction is removed, corresponding to the fifth row of Fig.~\ref{fig:2}. In the upper panel, we show instead the two-body density of the originally-recentered configurations. The effects are clearly visible and somewhat intuitive. The center-of-mass constraint produces dipole-type correlations, that is, a suppression for $\Delta\phi=0$ followed by a significant enhancement for $\Delta\phi=\pm\pi$. Comparing recentered vs non-recentered results, we see that correlations induced by recentering are in fact larger than those induced by the Fermi statistics.

In Fig.~\ref{fig:COM_UIX}, the same analysis is reported for the QMC configurations employing the AV6P+UIX potential. While the lower panel corresponds to the fourth row in Fig.~\ref{fig:2}, it is interesting to remark, in the upper panels, that the center-of-mass corrections acts against the $\alpha$ clustering correlations and the triangular deformation of the two-body density.

\subsection{Analytical results for independent fermions}

The correlation functions that we have studied in this paper can be obtained analytically for the mean-field model of $^{16}$O, corresponding to nucleons that move independently of each other in the mean field of a three-dimensional harmonic oscillator. In this picture, the ground state is obtained by filling the 1s and the three 1p states of the oscillator. The one-body density is then given by 
\[
    \rho^{(1)}(\mathbf{r})=4\frac{\alpha^3}{\pi^{3/2}} \text{e}^{-\alpha^2 \mathbf{r}^2} \left( 1+2\alpha^2 \mathbf{r}^2\right)
\] 
where $\alpha=\sqrt{m\omega/\hbar}$, with $m$ the nucleon mass and $\omega$ the frequency of the oscillator. The factor 4 accounts for spin and isospin. 

In a mean field approximation, the  connected part of the two-body density can be expressed in terms of the one-body density matrix $\rho(\mathbf{r}_1,  \mathbf{r}_2)$:
\[
 \rho^{(2)}_c(\mathbf{r}_1,\mathbf{r}_2) = -|\rho(\mathbf{r}_1,  \mathbf{r}_2)|^2 \,. 
\]
The density matrix can be written as $\rho(\mathbf{r}_1,  \mathbf{r}_2)=\sum_\lambda \varphi_\lambda(\mathbf{r}_1)\varphi_\lambda^*(\mathbf{r}_2)$, where the sum runs over all the occupied states $\lambda$, with $\varphi_\lambda(\mathbf{r})$ being the corresponding single particle wave functions. 
For the single-particle wave functions of the harmonic oscillator, the calculation is straightforward. One finds
\[
\rho(\mathbf{r}_1,  \mathbf{r}_2)=4\frac{\alpha^3}{\pi^{3/2}} \text{e}^{-\alpha^2 (\mathbf{r}^2+\mathbf{r}^2)/2} \left( 1+2\alpha^2 \mathbf{r}_1\cdot\mathbf{r}_2\right).
\]
Upon squaring and subsequently projecting on the transverse plane, one gets 
\begin{align}\label{eq:rhosqared}
\nonumber |\rho(\mathbf{x}_{1},  \mathbf{x}_{2})|^2&=\frac{2\alpha^4}{\pi^2}\text{e}^{-\alpha^2 (\mathbf{x}_{1}^2+\mathbf{x}_{2}^2}) \\ &\times\left[ 1+2\alpha^2 \mathbf{x}_{1}\cdot \mathbf{x}_{2} +2\alpha^4 (\mathbf{x}_{1}\cdot \mathbf{x}_{2})^2\right]
\end{align}
where we have set $ \mathbf{x}_{1}\equiv\mathbf{r}_{1\perp}$, $ \mathbf{x}_{2}\equiv\mathbf{r}_{2\perp}$. Setting $\mathbf{x}_{1}\cdot \mathbf{x}_{2}= x_1 x_2\cos\Delta\phi$, this expression leads to modulations proportional to  $\cos\Delta\phi$ and $\cos2\Delta\phi$ in the connected two-point function $\rho^{(2)}_c(\mathbf{r}_1,\mathbf{r}_2)$. The angular dependence originates from the three occupied p-wave orbitals, while the s-wave gives a contribution independent of the angle. This angular dependence is of quantum origin and a consequence of the Pauli exclusion principle. A classical approximation to $\rho^{(2)}_c(\mathbf{x}_1,\mathbf{x}_2)$ would yield $-(1/A) \rho^{(1)}(\mathbf{x}_1)\rho^{(1)}(\mathbf{x}_2)$ in place of $|\rho(\mathbf{x}_{1},  \mathbf{x}_{2})|^2$ in Eq.~(\ref{eq:rhosqared}).

For the quadrupole parameter in Eq.~(\ref{eq:normaEhat}) we get 
\begin{align}
 \langle \hat{R}_2\rangle &=\frac{1}{A}\int_{\mathbf{x}} \mathbf{x}^2\rho^{(1)}(\mathbf{x})=\frac{3}{2\alpha^2}\\
\nonumber \langle \hat{\mathcal{E}}_2\rangle &= \frac{1}{A(A-1)} \int_{\mathbf{x}_1\mathbf{x}_2} \rho_c^{(2)}(\mathbf{x}_1, \mathbf{x}_2) \, \mathbf{x}_1^2 \, \mathbf{x}_1^2 \, \text{e}^{2i  (\phi_1-\phi_2)}\\
 &= -\frac{1}{15 \alpha^4}.
\end{align} 
Choosing $\alpha=0.5435 \text{fm}^{-1}$ which reproduces the rms mean-field radius of $^{16}$O quoted in Table \ref{tab:1}, one gets the values $\langle \hat{R}_2\rangle=5.08 \,\text{fm}^2 $ and $\langle \hat{\mathcal{E}}_2\rangle=-0.764 \,\text{fm}^4$, and $\mathfrak{B}_2=0.249$.  There is no modulation proportional to $\cos 3\Delta\phi$, that is $\langle \hat{\mathcal{E}}_3 \rangle=0$. The values quoted in Table \ref{tab:1} are obtained with Woods-Saxon single particle wave functions and differ only slightly from those obtained with oscillator wave functions.

\end{document}